\documentclass[a4paper,twoside,10pt]{article}
\usepackage{graphicx,publaob}

\def\papertype{Contributed paper}
\begin{document}

\title{(NON)ARCHIMEDEAN QUANTUM COSMOLOGY AND TACHYONIC INFLATION}

\authors{G. S. DJORDJEVI\' C \lowercase{and} LJ. NE\v SI\' C}

\address{Department of Physics, University of Nis, P.O. Box 224, 18000 Nis, Serbia}
\Email{gorandj}{junis.ni.ac}{rs}

\markboth {{\scriptsize {\rm (NON)ARCHIMEDEAN QUANTUM COSMOLOGY AND
TACHYONIC INFLATION}}}{{\scriptsize {\rm G. S. DJORDJEVI\' C and LJ.
NE\v SI\' C}}}

\abstract{We review the relevance of quantum rolling tachyons and
corresponding inflation scenario in the frame of the standard,
$p$-adic and adelic minisuperspace quantum cosmology. The field
theory of tachyon matter proposed by Sen in a zero-dimensional
version suggested by Kar leads to a model of a particle moving in
a constant external field with quadratic damping. We calculate the
exact quantum propagator of the model, as well as, the vacuum
states and conditions necessary to construct an adelic
generalization. In addition we present an overview on several
important cosmological models on archimedean and nonarchimedean
spaces.}





\section{INTRODUCTION}

The main task of quantum cosmology (Wiltshire, 1996) is to
describe the evolution of the universe in the very early stage.
Usually one takes the universe is described by a complex-valued
wave function. Since quantum cosmology is related to the Planck
scale phenomena it is logical to consider various geometries (in
particular the nonarchimedean (Djordjevic et al., 2002) and
noncommutative (Garcia-Compean et al., 2002) ones) and
parametrization of the space-time coordinates: real, $p$-adic, or
even adelic (Vladimirov et al., 1994). In this article, we will
generally maintain space-time coordinates and matter fields to be
real and $p$-adic.

It is quite natural to consider that in the very early stage of
its evolution the universe is in a quantum state, which is
described by a wave function. Concerning the wave function, we
will here maintain the standard point of view: the wave function
takes complex values, but space-time coordinates and matter fields
will be treated in a more complete way to be adelic, i.e. they
have real as well as $p$-adic properties simultaneously. This
approach is motivated by the following reasons: ({\it i}) the
field of rational numbers $Q$, which contains all observational
and experimental numerical data, is a dense subfield not only in
the field of real numbers $R$ but also in the fields of $p$-adic
numbers $Q_p$ ($p$ is any prime number), ({\it ii}) there is a
plausible analysis (Vladimirov et al., 1994) within and over $Q_p$
as well as that one related to $R$, ({\it iii}) general
mathematical methods and fundamental physical laws should be
invariant under an interchange of the number fields $R$ and $Q_p$
(Volovich, 1987), ({\it iv}) there is a quantum gravity
uncertainty (Garay, 1995) $\Delta x$ while measuring distances
around the Planck length $\ell_0$,
\begin{equation}
 \Delta x \geq \ell_0 = \sqrt{\frac{\hbar G}{c3}} \sim 10^{-33} cm,
\label{1prim}
\end{equation}
which restricts priority of archimedean geometry based on real
numbers and gives rise to employment of nonarchimedean geometry
related to $p$-adic numbers (Volovich, 1987), ({\it v}) it seems
to be quite reasonable to extend compact  archimedean geometries
by the nonarchimedean ones in the path integral method, and ({\it
vi}) adelic quantum mechanics (Dragovich, 1994) applied to quantum
cosmology provides realization of all the above statements. The
successful application of $p$-adic numbers and adeles in modern
theoretical and mathematical physics started in 1987, in the
context of string amplitudes (Vladimirov et al., 1987; Freund and
Witten, 1987)(for a review, see Refs. (Freund and Witten, 1987;
Volovich et al., 1994; Khrennikov, 1997). For a systematic
research in this field it was formulated $p$-adic quantum
mechanics (Vladimirov and Volovich, 1989; Ruelle et al., 1989) and
adelic quantum mechanics (Dragovich, 1994; Dragovich, 1995). They
are quantum mechanics with complex-valued wave functions of
$p$-adic and adelic arguments, respectively. In the unified form,
adelic quantum mechanics contains ordinary and all $p$-adic
quantum mechanics. $p$-Adic gravity and the wave function of the
universe were considered in the paper (Aref'eva et al., 1991)
published in 1991.  An idea of the fluctuating number fields at
the Planck scale was introduced and it was suggested to restrict
the Hartle-Hawking (Hartle and Hawking, 1983) proposal to
summation only over algebraic manifolds. It was shown that the
wave function for the de Sitter minisuperspace model can be
treated in the form of an infinite product of $p$-adic
counterparts. Another approach to quantum cosmology, which takes
into account $p$-adic effects was proposed in 1995 (Dragovich,
1995). Like in adelic quantum mechanics, adelic eigenfunction of
the universe is a product of the corresponding eigenfunctions of
real and all $p$-adic cases. $p$-Adic wave functions are defined
by $p$-adic generalization of the Hartle-Hawking path integral
proposal. It was shown that in the framework of this procedure one
obtains an adelic wave function for the de Sitter minisuperspace
model. However, this procedure with the Hartle-Hawking $p$-adic
prescription does not work when matter fields are included into
consideration. The solution of this problem was found (Dragovich
and Nesic, 1999) in treating minisuperspace cosmological models as
models of adelic quantum mechanics.

Supernova Ia observations show that the expansion of the Universe
is accelerating (Perlmuter et al., 1999), contrary to
Friedmann-Robertson-Walker (FRW) cosmological models, with
non-relativistic matter and radiation. Also, cosmic microwave
background (CMB) radiation data are suggesting that the expansion
of our Universe seems to be in an accelerated state which is
referred to as the ``dark energy`` effect. The cosmological
constant as the vacuum energy can be responsible for this
evolution by providing negative pressure. A need for understanding
these new and rather surprising facts, including (cold) ``dark
matter``, has motivated numerous authors to reconsider different
inflation scenarios. Despite some evident problems (Sami et al.,
2004) such as insufficiently long period of inflation,
tachyon-driven scenarios (Gibbons, 2003; Choudhury and Ghoshal,
2002) or (Tranberg et al., 2007) remain highly interesting for
study.

There have been a number of attempts to understand this
description of the early Universe via (classical) nonlocal
cosmological models, first of all via $p$-adic inflation models
(Barnaby et al., 2007; Joukovskaya, 2007), which are represented
by nonlocal $p$-adic string theory coupled to gravity. For these
models, some rolling inflationary solutions were constructed and
compared with CMB observations. Another example is the
investigation of the $p$-adic inflation near a maximum of the
nonlocal potential when non-local derivative operators are
included in the inflaton Lagrangian. It was found that
higher-order derivative operators can support a (sufficiently)
prolonged phase of slow-roll inflation (Lidsey, 2007). These
results are an additional strong motivation for a rather general -
adelic approach to quantum cosmology.

In the unified form, adelic quantum mechanics contains ordinary
and all $p$-adic quantum mechanics. As there is not an appropriate
$p$-adic Schr\"odinger equation, there is also no $p$-adic
generalization of the Wheeler-De Witt equation. Instead of the
differential approach, Feynman's path integral method is exploited
(Djordjevic and Dragovich, 1997; Djordjevic and Dragovich, 2000;
Djordjevic et al., 1999; Dimitrijevic et al., 2008). $p$-Adic
gravity and the wave function of the universe were considered
(Aref'eva et al., 1991) as an idea of the fluctuating number
fields at the Planck scale. Like in adelic quantum mechanics, the
adelic eigenfunction of the universe is a product of the
corresponding eigenfunctions of real and all $p$-adic cases. It
was shown that in the framework of this procedure one obtains an
adelic wave function for the de Sitter minisuperspace model.
However, the adelic generalization with the Hartle-Hawking
$p$-adic prescription does not work well. Consideration has been
much more successful when minisuperspace cosmological models are
treated as models of adelic quantum mechanics. It is a strong
motivation to study a class of exactly solvable quantum mechanical
models and apply them in the frame of quantum cosmology. For the
review and detailed discussion see (Djordjevic et al., 2002;
Djordjevic et al., 2002a). The nonarchimedean and noncommutative
cosmological quantum models with extra dimensions and an
accelerating phase have been considered (Djordjevic and Nesic,
2005), as well as the relevant models and techniques in a pure
quantum mechanical context (Dimitrijevic et al., 2007; Djordjevic
and Nesic, 2005; Dimitrijevic et al., 2004; Djordjevic and Nesic,
2003). To keep this text to a reasonable size, we will not
consider noncommuative cosmology here, and a review would be given
elsewhere.

This review is organized as follows: after the Introduction, in
Chapter 2 we give basic information on ``$p$-adics`` and adeles.
Chapter 3 is devoted to $p$-Adic and Adelic quantum mechanics as
an underlaying formalism for the corresponding approach to quantum
cosmology briefly explained in the Chapter 4. Two simple
cosmological models in Chapter 5 illustrate application of
$p$-adic numbers in the Hartle-Hawking proposal, problems and
possible solutions for them we are faced with in this approach.
The most promising approach which generalize real and $p$ approach
to quantum cosmology is presented in Chapter 6, again through 2
simple examples. Chapter 7 and 8 are reserved for a quite
intriguing and hot problems in modern cosmology and, let us add,
string theory as well, namely inflation and tachyons. Following S.
Kar's  idea on the possibility of the examination of zero
dimensional theory of the field theory of (real) tachyon matter
(Kar, 2002), and motivated by successes and shortcomings of
classical $p$-adic inflation, we consider real and $p$-adic
aspects of a relevant model with quadratic damping. We calculated
the corresponding propagator and considered vacuum states for
$p$-adic and adelic tachyons. We end our paper with a short
conclusion and a few ideas for future research. A list of
references is quite subjective neither exhaustive nor complete
one, but could be useful for a reader interested in having a
better insight in results in application nonarchimedean geometry
in quantum theory and cosmology.

\section{$p$-ADIC NUMBERS AND ADELES}

We give here a brief survey of some basic properties of $p$-adic
numbers and adeles, which we exploit in this work. In addition, a
reach structure of nonarchimedean analysis and geometry, numerous
similarities as well as differences in respect to the ``standard``
ones will serve as a good basis for a general (re)consideration of
mathematical foundation od modern High Energy Physics.

Completion of $Q$ with respect to the standard absolute value
($|\cdot |_\infty$) gives $R$, and an algebraic extension of $R$
makes $C$. According to the Ostrowski theorem (Vladimirov et al.,
1994) any non-trivial norm on the field of rational numbers $Q$ is
equivalent to the absolute value $|\cdot|_\infty$ or to the
$p$-adic norm $|\cdot|_ p$, where $p$ is a prime number. $p$-Adic
norm is the nonarchimedean (ultrametric) one and for a rational
number, $0\ne x\in Q$, $\ x=p^\nu {m \over n}$, $\ 0\ne n,\nu,
m\in Z$, has a value $|x|_p=p^{- \nu}$. Completion of $Q$ with
respect to the $p$-adic norm for a fixed $p$ leads to the
corresponding field of $p$-adic numbers $Q_p$. Completions of $Q$
with respect to  $|\cdot |_\infty$ and all $|\cdot |_p$ exhaust
all possible completions of $Q$. $p$-Adic number $x\in Q_p$, in
the canonical form, is an infinite expansion
\begin{equation}
x=p^\nu\sum\limits_{i=0}\limits^\infty x_ip^i,\qquad x_0\ne0,\quad
0\leq x_i\leq p-1. \label{2.1}
\end{equation}
The norm of $p$-adic number $x$ in (\ref{2.1}) is $|x|_p=p^{-\nu}$
and satisfies not only the triangle inequality, but also the
stronger one
\begin{equation}
|x+y|_p\le\max (|x|_p,|y|_p). \label{2.2}
\end{equation}
Metric on $Q_p$ is defined by $d_p(x,y)=|x-y|_p$. This metric is the
nonarchimedean one and the pair ($Q_p,d_p$) presents locally
compact, topologically complete, separable and totally disconnected
$p$-adic metric space. In the metric space $Q_p$, $p$-adic ball
$B_\nu(a)$, with the centre at the point $a$ and the radius $p^\nu$
is the set
\begin{equation}
\label{2.3} B_\nu(a)=\{x\in Q_p:\ |x-a|_p\le p^\nu,\ \nu\in Z\}.
\label{2.4}
\end{equation}
$p$-Adic sphere $S_\nu(a)$ with the centre $a$ and the radius
$p^\nu$ is
\begin{equation}
\label{2.5} S_\nu(a)=\{x\in Q_p:\ |x-a|_p=p^\nu,\ \nu\in Z\}.
\end{equation}
Elementary $p$-adic functions (Schikhof, 2006) are given by the
series of the same form as in the real case, e.g.
\begin{equation}
\exp x=\sum_{k=0}^\infty \frac{x^k} {k!},
\qquad \sinh x=\sum_{k=0}^\infty \frac{x^{k+1}} {(2k+1)!}, \qquad
\cosh x=\sum_{k=0}^\infty \frac{x^{2k}} {(2k)!}, \label{2.8}
\end{equation}
\begin{equation}
\tanh x=\sum_{k=2}^\infty \frac{2^k(2^k-1)B_kx^{k-1}} {k!}, \qquad
\coth x= \frac{1} {x}+ \sum_{k=2}^\infty \frac{2^kB_kx^{k-1}} {k!},
\label{2.9}
\end{equation}
where $B_k$ are Bernoulli's numbers. These functions have the same
domain of convergence $G_p=\{ x\in Q_p : |x|_p < |2|_p \} $.


Real and $p$-adic numbers are unified in the form of the adeles
(Gel'fand et al., 1966). An adele is an infinite sequence
\begin{equation}
a=(a_\infty, a_2,...,a_p,...), \label{2.10}
\end{equation}
where $a_\infty\in Q_\infty$, and $a_p\in Q_p$, with restriction
to $a_p\in Z_p$ $(\ Z_p=\{x\in Q_p: |x|_p\leq1\})$ for all but a
finite set $S$ of primes $p$. If we introduce ${\cal
A}(S)=Q_\infty\times\prod\limits_{p\in S} Q_p\times\prod
\limits_{p \notin S} Z_p$ then the space of all adeles is
 ${\cal A}=\bigcup\limits_S{\cal A}(S)$, which  is a topological ring.
Namely, ${\cal A}$ is a ring with respect to the componentwise
addition and multiplication. A principal adele is a sequence
$(r,r,...,r,...)\in {\cal A}$, where $r\in Q$. Thus, the ring of
principal adeles, which is a subring of ${\cal A}$, is isomorphic to
$Q$. An important function on ${\cal A}$ is the additive character
$\chi(x), \ x\in {\cal A}$, which is a continuous and complex-valued
function with basic properties:
\begin{equation}
|\chi (x)|_\infty=1, \quad\chi(x +y)= \chi(x) \chi(y). \label{2.11}
\end{equation}
This additive character may be presented as
\begin{equation}
\chi(x)=\prod_\upsilon\chi_\upsilon(x_\upsilon)= \exp(-2\pi
ix_\infty)\prod_p \exp(2\pi i\{x_p\}_p), \label{2.12}
\end{equation}
where $ \upsilon=\infty,2,\cdots,p,\cdots$, and $\{x\}_p$ is the
fractional part of the $p$-adic number $x$. Map $\varphi : {\cal
A}\to C$, which has the form
\begin{equation}
\varphi(x) = \varphi_\infty(x_\infty)\prod_{p\in S}\varphi_p(x_p)
\prod_{p\not\in S}\Omega(\mid x_p\mid_p), \label{2.13}
\end{equation}
where $\varphi_\infty(x_\infty)\in D(Q_\infty)$ is an infinitely
differentiable function on $Q_\infty$ and falls to zero faster than
any power of $\mid x_\infty\mid_\infty$ as $\mid x_\infty\mid_\infty
\to\infty$,   $\ \varphi_p(x_p)\in D (Q_p)$ is a locally constant
function with compact support, and
\begin{equation}
\Omega (|x|_p) = \left\{\begin{array}{ll}
1, & |x|_p \leq 1,\\
0, & |x|_p >1,
\end{array}
\right. \label{2.14}
\end{equation}
is called an elementary function on ${\cal A}$. Finite linear
combinations of elementary functions (\ref{2.13}) make the set of
the Schwartz-Bruhat functions $D({\cal A})$. The existence of
$\Omega$-function is unavoidable for a construction of any adelic
model, in particular for quantum mechanical or quantum
cosmological model. The Fourier transform is
\begin{equation}
\tilde\varphi(\xi) = \int_{\cal A}\varphi(x)\chi(\xi x)dx
\label{2.15}
\end{equation}
and it maps one-to-one $D({\cal A})$ onto $D'({\cal A})$. It is
worth noting that $\Omega$-function is a counterpart of the
Gaussian in the real case, since it is invariant with respect to
the Fourier transform. It is also an important issue for
consideration of the ground state(s) of quantum mechanical systems
at high energies, where the use of $p$-adic numbers and
nonarchimedean geometry in ``modelling`` should be fully
justified.

The integrals of the Gauss type over the $p$-adic sphere $S_\nu$,
$p$-adic ball $B_\nu$ and over any $Q_v$ are (for $|4\alpha|_p\geq
p^{2-2\nu}$):
\begin{equation}
\label{2.16} \int_{S_\nu} \chi_p \left( \alpha x^2+\beta x
\right)dx = \left\{\begin{array}{ll}
\lambda_p(\alpha)|2\alpha|_p^{-1/2} \chi_p \left(
-\frac{\beta^2}{4\alpha} \right), &
\left|\frac{\beta}{2\alpha}\right|_p=p^\nu,\\
0, & \left| \frac{\beta}{2\alpha}\right|_p\neq p^\nu,
\end{array}
\right.
\end{equation}
\begin{equation}
\label{2.17} \int_{B_\nu}\chi_p(\alpha x^2+\beta x)dx =
\left\{\begin{array}{ll} p^{\nu}
\Omega(p^{\nu}|\beta|_p), & |\alpha|_pp^{2\nu}\leq1, \\
\frac{\lambda_p(\alpha)}
{|2\alpha|^{1/2}_p}\chi_p\left(-\frac{\beta^2}{4\alpha}\right)
\Omega\left(p^{-\nu}\left|\frac{\beta}{2\alpha}\right|_p\right),
 & |\alpha|_pp^{2\nu}>1,
\end{array}
\right.
\end{equation}
\begin{equation}
\label{2.18} \int_{Q_v}\chi_p(\alpha x^2+\beta x)dx=
\lambda_v(\alpha)|2\alpha |^{-1/2}_v
\chi_v\left(-\frac{\beta^2}{4\alpha}\right),\quad \alpha \ne 0.
\end{equation}
The  arithmetic functions $\lambda_v(a):\enskip  Q_v\mapsto C$,
where $v= \infty, 2, 3, 5,\cdots$, have the following properties:
\begin{equation}
|\lambda_v(a)|_\infty=1,\quad \lambda_v(0) =1, \quad
\lambda_v(ab^{2})=\lambda_v(a), \label{2.19}
\end{equation}
\begin{equation}
\lambda_v(a)\lambda_v(b)= \lambda_v(a+b)\lambda_v(a^{-1}+b^{-1}),
\label{2.20}
\end{equation}
where $a\ne0, \ b\ne0$. The physical implication of $\lambda_p$
function could be very important, but still far away to be well
understood. One of the most intriguing consequences would be
multidimensional nature of time (at Planck scale (Vladimirov et
al., 1994)).

\section{$p$-ADIC AND ADELIC QUANTUM MECHANICS}

In foundations of standard quantum mechanics (over $R$) one usually
starts with a representation of the canonical commutation relation
\begin{equation}
\label{3.1} [\hat q,\hat k]=i\hbar ,
\end{equation}
where $q$ is a spatial coordinate and $k$ is the corresponding
momentum. It is well-known that the procedure of quantization is
not unique. In formulation of $p$-adic quantum mechanics
(Vladimirov and Volovich, 1989; Ruelle et al., 1989) the
multiplication $\hat q\psi\rightarrow x \psi$ has no meaning for
$x\in Q_p$ and $\psi(x)\in C$. Also, there is no possibility to
define $p$-adic "momentum" or "Hamiltonian" operator. In the real
case they are infinitesimal generators of space and time
translations, but, since $Q_p$ is disconnected field, these
infinitesimal transformations become meaningless. However, finite
transformations remain meaningful and the corresponding Weyl and
evolution operators are $p$-adically well defined. Canonical
commutation relation in $p$-adic case can be represented by the
Weyl operators ($h=1$)
\begin{equation}
\hat Q_p(\alpha) \psi_p(x)=\chi_p(\alpha x)\psi_p(x) \label{3.2}
\end{equation}
\begin{equation}
\hat K_p(\beta)\psi(x)=\psi_p(x+\beta) . \label{3.3}
\end{equation}
Now, instead of the relation (\ref{3.1}) in the real case, we have
\begin{equation}
\hat Q_p(\alpha)\hat K_p(\beta)=\chi_p(\alpha\beta) \hat
K_p(\beta)\hat Q_p(\alpha) \label{3.4}
\end{equation}
in the $p$-adic one.

Dynamics of a $p$-adic quantum model is described by a unitary
operator of evolution $U(t)$ without using the Hamiltonian.
Instead of that, the evolution operator has been formulated in
terms of its kernel ${\cal K}_t(x,y)$
\begin{equation}
\label{3.7} U_p(t)\psi(x)=\int_{Q_p}{\cal K}_t(x,y)\psi(y) dy.
\end{equation}
In this way (Vladimirov and Volovich, 1989) $p$-adic quantum
mechanics is given by a triple
\begin{equation}
(L_2(Q_p), W_p(z_p), U_p(t_p)). \label{3.8}
\end{equation}
Keeping in mind that standard quantum mechanics can be also given
as the corresponding triple, ordinary and $p$-adic quantum
mechanics can be unified in the form of adelic quantum mechanics
(Dragovich, 1994; Dragovich, 1995)
\begin{equation}
(L_2({\cal A}), W(z), U(t)). \label{3.9}
\end{equation}
$L_{2}({\cal A})$ is the Hilbert space on ${\cal A}$, $W(z)$ is a
unitary representation of the Heisenberg-Weyl group on $L_2({\cal
A})$ and $U(t)$ is a unitary representation of the evolution
operator on $L_2({\cal A})$. The evolution operator $U(t)$ is
defined by
\begin{equation}
U(t)\psi(x)=\int_{{\cal A}} {\cal
K}_t(x,y)\psi(y)dy=\prod\limits_{v}{} \int_{Q_{v}}{\cal
K}_{t}^{(v)}(x_{v},y_{v})\psi^{(v)}(y_v) dy_{v}. \label{3.10}
\end{equation}
The eigenvalue problem for $U(t)$ reads
\begin{equation}
U(t)\psi _{\alpha \beta} (x)=\chi (E_{\alpha} t) \psi _{\alpha
\beta} (x), \label{3.11}
\end{equation}
where $\psi_{\alpha \beta}$ are adelic eigenfunctions, $E_{\alpha
}=(E_{\infty}, E_{2},..., E_{p},...)$ is the corresponding energy,
indices $\alpha$ and $\beta$ denote  energy levels and their
degeneration. Note that any adelic eigenfunction has the form
\begin{equation}
\label{3.12} \Psi(x) = \Psi_\infty(x_\infty)\prod_{p\in
S}\Psi_p(x_p) \prod_{p\not\in S}\Omega(\mid x_p\mid_p) , \quad x\in
{\cal A},
\end{equation}
where $\Psi_{\infty}\in L_2(R)$, $\Psi_{p}\in L_2(Q_p)$. A suitable
way to calculate $p$-adic propagator ${\cal K}_p (x'',t'';x',t')$ is
to use Feynman's path integral method, i.e.
\begin{equation}
{\cal K}(x'',t'';x',t') = \int_{x',t'}^{x'',t''} \chi_p \left(
-\frac{1}{h} \int_{t'}^{t''} L(\dot{q},q,t) dt  \right) {\cal D}q.
\label{3.13}
\end{equation}
It has been evaluated (Djordjevic and Dragovich, 1997; Djordjevic
et al., 2003) for quadratic Lagrangians in the same way for real
and $p$-adic case and the following exact general expression is
obtained:
\begin{equation}
\label{3.14} {\cal K}_v(x'',t'';x',t')= \lambda_v \left( -
\frac{1}{2h} \frac{\partial^2{\bar S}}{\partial x''\partial x'}
\right) \left| \frac{\partial^2{\bar S}}{\partial x''\partial x'}
\right|_v^{\frac{1}{2}} \chi_v(-\frac{1}{h} {\bar S}
(x'',t'';x',t')),
\end{equation}
where $\lambda_v$ functions satisfy relations (\ref{2.19}) and
(\ref{2.20}). When one has a system with more than one dimension
with uncoupled spatial coordinates, then the total propagator is
the product of the corresponding one-dimensional propagators. As
an illustration of $p$-adic and adelic quantum-mechanical models,
the following one-dimensional systems with the quadratic
Lagrangians were considered: a free particle and harmonic
oscillator (Vladimirov et al., 1994; Dragovich, 1994; Dragovich,
1995), a particle in a constant field, a free relativistic
particle (Djordjevic et al., 1999) and a harmonic oscillator with
time-dependent frequency (Djordjevic and Dragovich, 2000). Adelic
quantum mechanics takes into account ordinary as well as $p$-adic
quantum effects and may be regarded as a starting point for
construction of a more complete superstring and M-theory. In the
low-energy limit adelic quantum mechanics becomes the ordinary one
(Djordjevic et al., 1999).

\section{QUANTUM COSMOLOGY}

According to the so-called standard cosmological model, in the
very beginning the universe was very small, dense, hot and started
to expand very fast. This initial period of evolution should be
unavoidably described by quantum cosmology. In the path integral
approach to quantum cosmology over the field of real numbers $R$,
the starting point is the idea that the amplitude to go from one
state with intrinsic metric $h_{ij}^\prime$, and matter
configuration $\phi^\prime$ on an initial hypersurface
$\Sigma^\prime$, to another state with metric
$h_{ij}^{\prime\prime}$, and matter configuration
$\phi^{\prime\prime}$ on a final hypersurface
$\Sigma^{\prime\prime}$, is given by a functional integral of the
form
\begin{equation}
\langle h_{ij}'',\phi'',\Sigma''| h_{ij}',\phi',\Sigma'\rangle =
\int {\cal D}{g_{\mu\nu}} {\cal D}\Phi e^{-S[g_{\mu\nu},\Phi]},
\label {4.1}
\end{equation}
over all four-geometries $g_{\mu\nu}$, and matter configurations
$\Phi$, which interpolate between the initial and final
configurations. In this expression $S[g_{\mu\nu},\Phi]$ is an
Einstein-Hilbert action for the gravitational and matter fields
(which can be massless, minimally or conformally coupled with
gravity). This expression stays valid in the $p$-adic case too,
because of its form invariance under change of real to the $p$-adic
number fields.

Among many cosmological models, there is one very important type
of models, the so-called de Sitter models. The de Sitter are the
models with the cosmological constant $\Lambda$ and without matter
fields. Models of this type are exactly soluble models and because
of that, they play a role similar to a linear harmonic oscillator
in ordinary quantum mechanics. The corresponding Einstein-Hilbert
action is (Halliwell and Myers, 1989)
\begin{equation}
\label{5.2} S= \frac{1}{16\pi G}\int_M d^Dx{\sqrt
{-g}}(R-2\Lambda) + \frac{1}{8\pi G}\int_{\partial M}d^{D-1}x\sqrt
h K,
\end{equation}
where R is the scalar curvature of $D$-manifold $M$, $K$ is the
trace of the extrinsic curvature $K_{ij}$ of the boundary
$\partial M$ of the $D$-manifold $M$. The general form of the
metric for these models is
\begin{equation}
\label{5.3} ds^2=\sigma^2 [-N^2dt^2+a^2(t)d\Omega^2_{D-1}],
\end{equation}
where $d\Omega^2_{D-1}$ denotes the metric on the unit
$(D-1)$-sphere

$$\sigma^{D-2}=\frac{{8\pi G}}{{V^{D-1}(D-1)(D-2)}}$$,

\noindent and $\ V^{D-1}$ is the volume of the unit
$(D-1)$-sphere. In the $D=3$ case, this model is related to the
multiple sphere configuration and wormhole solutions.
$\upsilon$-Adic ($\upsilon =\infty$ for the real, and $\upsilon =
p$ in the $p$-adic cases) classical action for this model is
\begin{equation}
\label{5.4} \bar S_\upsilon(a'',N; a',0)= \frac{1}{2\sqrt\lambda}
\left[ N\sqrt\lambda + \lambda \left(
\frac{2a''a'}{\sinh(N\sqrt\lambda)} -
\frac{a'^2+a''^2}{\tanh(N\sqrt\lambda)}\right)\right].
\end{equation}
Let us note that $a$ denotes a scale factor and $\lambda$ denotes
here the appropriately rescaled cosmological constant $\Lambda$,
i.e. $\lambda=\sigma^2\Lambda$. This model was investigated in all
aspects ($p$-adic, real and adelic) in Ref. (Djordjevic et al.,
2002). Especially, for this model, the adelic wave function (which
unifies the wave function over the field of real numbers and wave
functions over the field of $p$-adic numbers), is in the form
\begin{equation}
\label{5.5prim} \Psi(a)=\Psi_\infty(a)\prod_{p}\Psi_p(a_p),
\end{equation}
where $\Psi_\infty(a)$ is a standard wave function and
$\Psi_p(a_p)$ are $p$-adic wave functions. It is very important
that only for finite numbers of $p$, $p$-adic wave functions can
be different from $\Omega$ function which is defined by the
(\ref{2.14}).

At this place we indicate a considerable similarity between the
action (\ref{5.4}) for the de Sitter model in 2+1 dimensions and
the action (\ref{lj3}) for the tachyon field in the zero
dimensional model, i.e. ``quadratically damped particle under
gravity``. Tachyon fields, inflation and their classical and
quantum aspects are discussed in Chapters 8 and 9.

\section{$p$-ADIC MODELS IN THE HARTLE-HAWKING PROPOSAL}

The Hartle-Hawking proposal for the wave function of the universe
is generalized to $p$-adic case in Refs. (Dragovich and Nesic,
1996; Dragovich, 1995). In this approach, $p$-adic wave function
is a solution of the integral
\begin{equation}
\label{5.1} \Psi_p(q^\alpha)=\int_{|N|_p\leq1}dN{\cal
K}_p(q^\alpha,N;0,0),
\end{equation}
where $p$-adic integration has to be performed over the $p$-adic
ball $B_0$.

\subsection{MODELS OF THE DE SITTER TYPE}

It is well-known that there are many, many cosmological models. It
happens, surprisingly or not, that one of the very first models,
the de Sitter model, despite its simplicity, and at the first
glance artificiality, is still very important and actual one.
There are also many variations of the ``original`` de Sitter
model, all of them named as models of ``de Sitter type``. As it is
mentioned and well-known, models of the de Sitter type are models
with cosmological constant $\Lambda$ and without matter fields. We
consider two minisuperspace models of this type, with $D=4$ and
$D=3$ space-time dimensions. For $D=3$ using (\ref{3.14}) for the
propagator of this model we have
\begin{equation}
\label{5.5} {\cal K}_\upsilon(a'',N;a',0) = \lambda_\upsilon
\left(-\frac{2\sqrt\lambda}{\sinh (N\sqrt\lambda)} \right) \left|
\frac{\sqrt\lambda}{\sinh(N\sqrt\lambda)} \right|_\upsilon^{1/2}
\chi_\upsilon(-\bar S_\upsilon(a'',N;a',0)).
\end{equation}
The $p$-adic Hartle-Hawking wave function is
\begin{equation}
\label{5.6} \Psi_p(a,\lambda)= \int_{|N|_p\leq 1}
dN\frac{\lambda_p(-2N)}{|N|_p^{1/2}} \chi_p \left(
\frac{\sqrt\lambda\coth(N\sqrt\lambda)}{2}a^2 \right),
\end{equation}
which after $p$-adic integration becomes
\begin{equation}
\label{5.7} \Psi_p (a,\lambda) = \left\{\begin{array}{ll}
\Omega(|a|_p), & |\lambda|_p\leq p^{-2}\\
\frac{1}{2}\Omega(|a|_2), & |\lambda|_2\leq 2^{-4}.
\end{array}
\right.
\end{equation}
The de Sitter model in $D=4$ space-time dimensions is described by
the metric (Halliwell and Luoko, 1989)
\begin{equation}
\label{5.8} ds^2= \sigma^2 \left(
-\frac{N^2}{q(t)}dt^2+q(t)d\Omega_3^2 \right).
\end{equation}
For the $\upsilon$-adic  classical action
\begin{equation}
\label{5.9} \bar S_\upsilon(q'',T;q',0) = \frac{\lambda^2T^3}{24}
- [\lambda(q'+q'')-2 ]\frac{T}{4} - \frac{(q''-q')^2}{8T}
\end{equation}
the corresponding propagator is
\begin{equation}\label{5.10}
{\cal K}_\upsilon(q'',T|q',0)=
\frac{\lambda_\upsilon(-8T)}{|4T|_\upsilon^{1/2}}
\chi_\upsilon(-\bar S_\upsilon(q'',T|q',0)).
\end{equation}



\subsection{MODEL WITH A HOMOGENEOUS SCALAR FIELD}

To deal with the models of the de Sitter type  is very
instructive. However, it is also important to consider models with
some matter content. If we use metric in the form (Garray et al.,
1991)
\begin{equation}
\label{5.12} ds^2=
\sigma^2\left(-N^2(t)\frac{dt^2}{a^2(t)}+a^2(t)d\Omega^2_3\right),
\end{equation}
the gravitational part of the action in the form (\ref{5.2}) (with
$D=4$), and a suitable action for a scalar field
\begin{equation}
\label{5.13} S_{matter}= -\frac{1}{2}\int_M d^4x\sqrt{-g} \left[
g^{\mu\nu}\partial_\mu\Phi\partial_\nu\Phi+V(\Phi) \right],
\end{equation}
then, after some substitutions, we get the corresponding classical
action and propagator as follows:
$$
\bar S_p(x'',y'',N|x',y',0)= \frac{\alpha^2-\beta^2}{24}N^3+
\frac{1}{4}(2-\alpha(x'+x'')-\beta(y'+y''))N
$$
\begin{equation}
\label{5.14} +\frac{-(x''-x')^2+(y''-y')^2}{8N},
\end{equation}
\begin{equation}
{\cal K}_p(x'',y'',N|x',y',0) =\frac{1}{|4N|_p} \chi_p(-\bar
S_p(x'',y'',N;x',y',0)).
\end{equation}
As we have shown (Dragovich and Nesic, 1999) for this model, a
$p$-adic Hartle-Hawking wave function in the form of $\Omega$ -
function does not exist. This leads to the conclusion that either
the above model is not adelic, or that $p$-adic generalization of
the Hartle-Hawking proposal is not an adequate one. However, if in
the action (\ref{5.14}) we take $\beta=0,\ y=0$, then we get
classical action for the de Sitter model (\ref{5.9}), and such
model, as we showed it, is the adelic one. The similar conclusion
holds also for some other models in which minisuperspace is not
one-dimensional. This is a reason to regard $p$-adic and adelic
minisuperspace quantum cosmology just as the corresponding
application of $p$-adic and adelic quantum mechanics without the
Hartle-Hawking proposal.

We demonstrated that $p$-adic (and adelic) to the de Sitter
model(s) works quite well, at least at a ``formal`` level, and can
be used in higher dimensional spaces. These results can be useful
for examination of the accelerating scenarios of an expanding
Universe.

\section{MINISUPERSPACE MODELS IN $p$-ADIC \break
{\hbox{\hskip.5cm}} AND ADELIC QUANTUM MECHANICS}

In this approach we investigate conditions under which
quantum-mechanical $p$-adic ground state exists in the form of
$\Omega$-function and some other eigenfunctions. This approach
leads to the desired result and it enables adelization of all
exactly soluble minisuperspace cosmological models, usually with
some restrictions on the parameters of the models. One can
speculate, but also continue a study, that nonarchimedean geometry
or ``nonarchimedean phase`` in evolution of the Universe restricts
a set of initial conditions and a set of Lagrangians related to a
realistic dynamics of our Universe (Djordjevic et al. 2000). The
necessary condition for the existence of an adelic model is an
existence of $p$-adic quantum-mechanical ground state
$\Omega(|q_\alpha|_p)$, i.e.
\begin{equation}
\label{1} \int_{|{q_\alpha}'|_p\leq1}{\cal K}_p
({q_\alpha}'',N;{q_\alpha}',0)d{q_\alpha}'=
\Omega(|{q_\alpha}''|_p),
\end{equation}
and, analogously, if a system  is in the state
$\Omega(p^\nu|q_\alpha|_p)$. If $p$-adic  ground state is of the
form of the $\delta$-function, we will investigate conditions
under which the corresponding kernel of the model  satisfies
equation
\begin{equation}
\label{3} \int_{Q_p}{\cal
K}_p(q_\alpha'',T;q',0)\delta(p^\nu-|q_\alpha'|_p)dq_\alpha'=
\chi_p(ET)\delta(p^\nu-|q_\alpha''|_p),
\end{equation}
with zero energy $E=0$. In the following, we apply (\ref{1}) and
(\ref{3}) to the some minisuperspace models.

\subsection{THE DE SITTER MODEL IN $D=3$ DIMENSIONS}

Let us demonstrate quantum mechanical approach to $p$-adic (real
and adelic as well) cosmological models through two simple but
relevant and instructive examples. In case we choose the de Sitter
model in 3 dimensions, application of the above exposed formalism
of $p$-adic quantum mechanics enable us to calculate (Dragovich
and Nesic, 1999) the ground state of the Universe
\begin{equation}
\label {6.4} \Psi_p(a,N) = \left\{\begin{array}{lr}
\Omega(|a|_p), & |N|_p\leq 1,\enskip  p\neq 2, \\
\Omega(|a|_2), & |N|_2\leq\frac{1}{4},\enskip p=2,
\end{array}
\right.
\end{equation}
with conditions $|\lambda|_p\leq1$ and $|\lambda|_2\leq 2$. We also
found
\begin{equation}
\label{6.5} \Psi_p(a,N)= \left\{\begin{array}{lr}
\Omega(p^\nu|a|_p), & |N|_p\leq p^{-2\nu},\qquad|\lambda|_p\leq p^{4\nu} \\
\Omega(2^\nu|a|_2), & |N|_2\leq 2^{-2-2\nu},|\lambda|_2\leq
2^{1+4\nu},
\end{array}
\right.
\end{equation}
where $\nu=1,2,\dots$\ . The existence of the ground state in the
form of the $\delta$-function may be investigated by the equation
(\ref{3}), i.e.
\begin{equation}
\label{6.6} \int_{Q_p}{\cal K}(a'',N;a',0)\delta(p^\nu-|a'|)da'=
\delta(p^\nu-|a''|),
\end{equation}
with the kernel (\ref{5.5}), what leads to the equation
$$
\lambda_p \left( -\frac{\sqrt\lambda}{2\sinh(N\sqrt\lambda)}
\right) \left| \frac{\sqrt\lambda}{\sinh(N\sqrt\lambda)}
\right|_p^{1/2}\chi_p \left(
-\frac{N}{2}+\frac{\sqrt\lambda}{2\tanh(N\sqrt\lambda)}{a''}^2
\right)
$$
\begin{equation}
\label{6.7} \times \int_{|a'|_p=p^\nu}\chi_p \left(
\frac{\sqrt\lambda}{2\tanh(N\sqrt\lambda)}{a'}^2
-\frac{\sqrt\lambda}{\sinh(N\sqrt\lambda)}{a''}a' \right)da'
=\delta(p^\nu- |a''|_p).
\end{equation}
The above integration is performed over $p$-adic sphere
 with the radius $p^\nu$ and for
$|\frac{N}{2}|_p\le p^{2\nu-2} $, $\nu=1,0,-1,\dots$\ . As a result,
on the left hand side we have
$$ \chi_p \left(
-\frac{N}{2}+\frac{\sqrt\lambda}{2}\tanh(N\sqrt\lambda){a''}^2
\right). $$ To have an equality, the norm of argument of the
additive character must be equal or less than unity. This
requirement leads to the condition $$ \left|
\frac{\sqrt\lambda\tanh(N\sqrt\lambda){a''}^2}{2}
\right|_p=p^{4\nu-2}|\lambda|_p\leq 1,\enskip\Leftrightarrow
\enskip|\lambda|_p\leq p^{2-4\nu}. $$ This  (for the $p$-adic
norms of $N$ and $\lambda$) is also related to the domain of
convergence of the analytic function $\tanh x$, $$
|N\sqrt\lambda|_p\leq|N|_p|\lambda|_p^{1/2}\leq p^{2\nu-2}\cdot
p^{1-2\nu} =p^{-1},\ \forall\nu. $$ If $p=2$, then condition
$|N|_2\leq 2^{2\nu-3}$ holds, for $\nu=1,0,-1,-2,\cdots$, and we
are in the domain of convergence. Finally, we conclude that also
$p$-adic ground state
\begin{equation}
\label{6.8} \Psi_p(a,N)= \left\{\begin{array}{rl}
\delta(p^\nu-|a|_p), & |N|_p\leq p^{2\nu-2},\quad|\lambda|_p\leq p^{2-4\nu},\\
\delta(2^\nu-|a|_2), & |N|_2\leq2^{2\nu-3},\quad|\lambda|_2\leq
2^{-4\nu},
\end{array}
\right.
\end{equation}
exists for $\nu=1,0,-1,-2,\dots$.

We can note that in this approach there is $\Omega$ function as a
ground state - wave function of the Universe. In fact, there is
more than one state we can consider as the vacuum state. It also
means that structure of $p$-adic (adelic) vacuum is quite rich and
can be a source of several scenarios in the inflation theory,
including tachyons.

\subsection{MODEL WITH A HOMOGENEOUS SCALAR FIELD}

When one considers the two-dimensional minisuperspace model with
two decoupled degrees of freedom (introduced in Subsection 5.2) we
find that the corresponding ground state is of the form
$\Omega(|x|_p)\Omega(|y|_p)$, i.e.
\begin{equation}
\label{6.13} \Psi_p(x,y,N)= \left\{\begin{array}{rl}
\Omega(|x|_p)\Omega(|y|_p), & |N|_p\leq1,\\
\Omega(|x|_2)\Omega(|y|_2), & |N|_2\leq\frac{1}{2},
\end{array}
\right.
\end{equation}
with $\alpha=4\cdot3\cdot l_1$, $\ \beta=4\cdot3\cdot l_2$, $\
l_1,l_2\in Z$, and also
\begin{equation}
\label{6.14} \Psi_p(x,y,N)= \left\{\begin{array}{rl}
\Omega(p^\nu|x|_p)\Omega(p^\mu|y|_p), & |\alpha|_p\leq
|3|_p^{1/2}p^{3\nu},
\ |\beta|_p\leq |3|_p^{1/2}p^{3\mu},\\
\Omega(2^\nu|x|_2)\Omega(2^\mu|y|_2), & |\alpha|_2\leq
2^{3\nu-1},\quad |\beta|_2\leq 2^{3\mu-1},
\end{array}
\right.
\end{equation}
where $\nu,\mu=1,2,3,...$. As in the previous cases, we can also
investigate the existence of the vacuum state of the form
$\delta(p^\nu-|x|_p)\delta(p^\nu-|y|_p).$ After some calculations
we find $p$-adic wave function for the ground state to be in the
form
\begin{equation}
\Psi_p(x,y,N) = \label{6.15} \left\{\begin{array}{rl}
\delta(p^\nu-|x|_p)\delta(p^\mu-|y|_p), &
     |N|_p\leq p^{2\nu,\mu-2},\ |\alpha,\beta|_p\leq p^{2-3\nu,\mu},\\
\delta(2^\nu-|x|_2)\delta(2^\mu-|y|_2), &
     |N|_2\leq2^{2\nu,\mu-1},\ |\alpha,\beta|_2\leq 2^{-3\nu,\mu},
\end{array}
\right.
\end{equation}
where $\nu,\mu=0,-1,-2,\dots$.

It means that the ``quantum mechanical`` approach to quantum
cosmology does not have an obvious contradiction and instability.
As we know, in the $p$-adic generalization of Hartle-Hawking
approach, for the multidimensional minisuperspace, a ground state
(in particular $\Omega$ function) is missing.



\section{$p$-ADIC INFLATION}

Cosmological inflation has become an integral part of the standard
model of the universe. It provides important clues for structure
formation in the universe and is capable of removing the
shortcomings of standard cosmology.

Many string theorists and cosmologists have turned their attention
to building and testing stringy models of inflation in recent
years. The goals have been to find natural realizations of
inflation within string theory, and novel features which would
help to distinguish the string-based models from their more
conventional field theory counterparts. In most examples to date,
string theory has been used to derive an effective 4D field theory
operating at energies below the string scale and all the
inflationary predictions are made within the context of this low
energy effective field theory. This is a perfectly valid approach
to string cosmology, but, at least, a few problems still exist.
For instance it is often very difficult to identify features of
string theory inflation that cannot be reproduced in more
conventional models. Thus, there is motivation to consider models
in which inflation takes place at higher energy scales where
stringy corrections to the low energy effective action are playing
an important role. This is usually daunting since the field theory
description should be supplemented by an infinite number of higher
dimensional operators at energies above the string scale, whose
detailed form is not known.  Because of that, to study nonlocality
(intimately connected with $p$-adic and nonarchimedean themes
(Dragovich, 2009), as ubiquitous in string field theory, and to
consider a broad class of nonlocal inflationary models is a quite
interesting area of research.

Gibbons (Gibbons, 2003) has emphasized the cosmological
implication of tachyonic condensate rolling towards its ground
state. The tachyonic matter might provide an explanation for
inflation at the early epochs and could contribute to a new form
of dark matter at later times. A recent paper on $p$-adic
inflation (Barnaby et al., 2007) gives rise to the hopes that
nonlocal inflation can succeed where the real string theory fails.
$p$-Adic string theory, initiated by Volovich and his pioneering
paper (Volovich, 1987) and developed by Arefeva, Dragovich,
Goshal, Frampton, Freund, Sen, Witten and many other, despite some
open and serious problems is an interesting and wide field of
research. For a review and a considerable list of references see
(Dragovich et al., 2009). In our approach we will just use some
results relevant for real and $p$-adic tachyons, suitable for
further study in inflation and suggest to a motivated reader to
find more details in the noted references.

 Starting from the action
of the $p$-adic string, with $m_s$ the string mass scale and $g_s$
the open string coupling constant,

\begin{equation}
\label{stringaction} S=\frac{m_s^4}{g_p^2}\int d^4 x\left(
-\frac{1}{2}\phi
p^{-\frac{-\partial_t^2+\bigtriangledown^2}{2m_s^2}}\phi+\frac{1}{p+1}\phi
^{p+1} \right), \, \, \,
\frac{1}{g_p^2}=\frac{1}{g_s^2}\frac{p^2}{p-1},
\end{equation}
for the open string tachyon scalar field $\phi (x)$, it has been
shown that a $p$-adic tachyon drives a sufficiently long period of
inflation while rolling away from the maximum of its potential.
Even though this result is constrained by $p\gg 1$ and obtained by
an approximation, it is a good motivation to consider $p$-adic
inflation for different tachyonic potentials. In particular, it
would be interesting to study $p$-adic inflation in quantum regime
and in adelic framework to overcome the constraint $p\gg 1$, with
an unclear physical meaning. For more details, and further
development see (Joukovskaya 2007).

\section{CLASSICAL AND QUANTUM TACHYONS}

A. Sen proposed a field theory of tachyon matter a few years ago
(Sen, 2002, 2005). The action is given as:
\begin{equation}
  \label{eq:bpu1}
  S=-\int d^{D+1} x V(T) \sqrt{1+\eta^{ij}\partial_i T\partial_j T}
\end{equation}
where $\eta_{00}=-1$ and
$\eta_{\alpha\beta}=\delta_{\alpha\beta}$, $\alpha,
\beta=1,2,3,...,D$, $T(x)$ is the scalar tachyon field and $V(T)$
is the tachyon potential which unusually appears in the action as
a multiplicative factor and has (from string field theory
arguments) exponential dependence with respect to the tachyon
field $V(T)\sim e^{-\alpha T/2}$. In this paper we will focus our
attention on this type of the potential. It is very useful to
understand and to investigate lower dimensional analogs of this
tachyon field theory. The corresponding zero dimensional analog of
a tachyon field can be obtained by the correspondence:
$x^i\rightarrow t, T\rightarrow x, V(T)\rightarrow V(x)$. The
action reads
\begin{equation}
  \label{eq:bpu3}
  S=-\int dtV(x) \sqrt{1-\dot x^2}.
\end{equation}
In what follows, all variables and parameters can be treated as real
or $p$-adic without a formal change in the obtained forms. It is not
difficult to see that action (\ref{eq:bpu3}), with some appropriate
replacement leads to the equation of motion for a particle with mass
$m$, under a constant external force, in the presence of quadratic
damping:
\begin{equation}
  \label{eq:bpu5}
  m\ddot y + \beta \dot y^2 = mg.
\end{equation}
This equation of motion can be obtained from two Lagrangians (Jain
et al., 2007; Dimitrijevic et al., 2008):
\begin{equation}
  \label{eq:eq3}
  L(y,\dot y)=\left (\frac{1}{2}m\dot y^2 +
\frac{m^2g}{2\beta}\right )e^{2{\frac{\beta} {m}}
  y},
  \end{equation}
\begin{equation}
  \label{eq:eq4}
  L(y, \dot y)=-e^{-\frac{\beta}{m} y}\sqrt{1-\frac{\beta}{mg}\dot
  y^2}.
\end{equation}
Despite the fact that different Lagrangians can give rise to
nonequivalent quantization, we will choose the form (\ref{eq:eq3})
that can be handled easily. The first one is better because of the
presence of the square root in the second one. The general solution
of the equation of motion is
\begin{equation} \label{gensol}
y(t)=C_2+\frac{m}{\beta}\ln[\cosh(\sqrt{\frac{g\beta}{m}}\,t+C_1)].
\end{equation}
For the initial and final conditions $y'=y(0)$ and $y''=y(T)$, for
the $\upsilon$-adic classical action we obtain
\begin{equation}
  \label{classact}
\bar
S_\upsilon(y'',T;y',0)=\frac{\sqrt{mg\beta}}{2\sinh(\sqrt{\frac{g\beta}{m}
} T)}\left[(e^{\frac{2\beta}{m}y'}+e^{\frac{2\beta}{m}y''})\cosh
(\sqrt{\frac{g\beta}{m}} T)-2e^{\frac{\beta}{m}(y'+y'')}\right].
\end{equation}

In the $p$-adic case, we get a constraint which arises from the
investigation of the domain of a convergence analytical function
which appears during the derivation of the formulae (\ref{gensol}).
This constraint is $|\dot y|_p\leq \frac{1}{p}|\sqrt{\frac{g
m}{\beta}}|_p$.

By the transformation $X=\frac{m}{\beta}e^{\frac{\beta} {m}y}$, we
can convert Lagrangian (\ref{eq:eq3}) in a more suitable, quadratic
form
\begin{equation}
  \label{eq:eq6}
  L(X,\dot X)=\frac{m\dot X^2}{2} +\frac{g\beta X^2}{2}.
\end{equation}

For the conditions $X'=X(0)$, and $X''=X(T)$, action for the
classical $\upsilon$-adic solution $\bar X(t)$ is
\begin{equation}
  \label{lj3}
\bar S_\upsilon(X'',T;X',0)= \frac { \sqrt{mg\beta} } { 2\sinh(\sqrt
{\frac{g\beta}{ m}}T)} \left[(X'^2+X''^2) \cosh(\sqrt{\frac{g\beta}{
m}} T)-2X'X''\right].
\end{equation}
We note that this action is different from the action (\ref{5.4})
only in one constant term. Because action (\ref{lj3}) is quadratic
one (with respect to the initial and final point), the
corresponding kernel is (Djordjevic and Dragovich 1997)
\begin{equation}
  \label{lj4}
  {\cal K}_\upsilon
(X'',T;X',0)= \lambda_\upsilon \left(\frac{1}{2h} \frac{\sqrt{g\beta
m}} {\sinh(\sqrt{\frac{g\beta}{ m}}
  T)}\right)
  \left|\frac{1}{h}
  \frac{\sqrt{g\beta m}}{\sinh(\sqrt{ \frac{g\beta}{ m}}
  T)}
  \right|_\upsilon^{1/2}
  \chi_\upsilon\left(-\frac{1}{h}\bar
S_\upsilon\right),
\end{equation}
where $\chi_\upsilon$ is the $\upsilon$-adic additive character
(Vladimirov 1994).

In what follows, we apply (\ref{1}) and (\ref{3}) to our model. As
a result for the $p$-adic wave functions (in the case $p\neq 2$),
we get
\begin{equation}
\label{wf1} \Psi_p(X)=\Omega(|X|_p), \quad |T|_p\leq
\left|\frac{m}{2h}\right|_p,\quad \left|\frac{g\beta
m}{4h^2}\right|_p<1
\end{equation}
\begin{equation}
\label{wf2} \Psi_p(X)=\Omega(p^\nu|X|_p), \quad |T|_p\leq
\left|\frac{m}{2h}\right|_p p^{-2\nu},\quad \left|\frac{g\beta
m}{4h^2}\right|_p\leq p^{3\nu}
\end{equation}
\begin{equation}
\label{wf3} \Psi_p(X)=\delta(p^\nu-|X|_p), \quad
\left|\frac{T}{2}\right|_p\leq \left|\frac{m}{h}\right|_p
p^{2\nu-2},\quad \left|\frac{g\beta m}{h^2}\right|_p\leq
p^{2-3\nu}.
\end{equation}

The above conditions are in accordance with the conditions for the
convergence of the $p$-adic analytical functions which appear in the
solution of the equation of motion (\ref{gensol}) and the classical
action (\ref{classact}). We see there is a wide freedom in choosing
the parameters of the model, such as mass of the tachyon field $m$,
damping factor $\beta$, parameter $g$ related to the ``strength of
the constant gravity``, and cosmological constant $\Lambda$ which
appears in the de Sitter $(2+1)$ dimensional model. A relevant
physical conclusion served from these relations still needs a more
realistic model with tachyon matter and with a precise form of
metrics.

\section{CONCLUSION}

In this paper, we find  applications of $p$-adic numbers in quantum
cosmology  very interesting. It gives new possibilities to
investigate the structure of space-time at the Planck scale. In the
Hartle-Hawking approach the wave function of a spatially closed
universe is defined by Feynman's path integral method. The action is
a function of the gravitational and matter fields, and integration
is performed over all compact real four-metrics connecting two
three-space states. According to Feynman's integration over all real
compact metrics, this approach generalizes to all corresponding
compact $p$-adic metrics. However, it does not lead to the adequate
adelic picture and generalization for a wide class of the
minisuperspace models. From the other side, the consideration of
minisuperspace models in the framework of adelic quantum mechanics
gives the appropriate
 adelic generalization. Moreover, we can conclude that all
these models lead to the picture of space-time as a discrete one.
Namely, for all the above models there exists adelic wave function
\begin{equation}
\label{7.1}
\Psi(q^1,...,q^n)=\prod_{\alpha=1}^n\Psi_\infty(q^\alpha_\infty)
\prod_p\prod_{\alpha=1}^n\Omega(|q^\alpha_p|_p),
\end{equation}
\noindent where $\Psi_\infty(q^\alpha_\infty)$ are the corresponding
wave functions of the universe in standard cosmology. Adopting the
usual probability interpretation of the wave function (\ref{7.1}) in
rational points of $q^\alpha$, we have
\begin{equation}
\label{7.2} \left| \Psi(q^1,...,q^n) \right|_\infty^2 =
\prod_{\alpha=1}^n \left| \Psi_\infty(q^\alpha) \right|_\infty^2
\prod_p\prod_{\alpha=1}^n\Omega(|q^\alpha|_p),
\end{equation}
\noindent because $(\Omega(|q^\alpha|_p))^2=\Omega(|q^\alpha|_p)$.
As a consequence of $\Omega$-function properties we have
\begin{equation}
\label{7.3} \left| \Psi(q^1,\dots,q^n) \right|_\infty^2= \cases
{\left|\Psi_\infty(q^\alpha)\right|_\infty^2, &$q^\alpha\in Z$,\cr
0, &$q^\alpha\in Q\backslash Z$.\cr}
\end{equation}
This result leads to the discretization of minisuperspace
coordinates $q^\alpha$, because probability is nonzero only in the
integer points of $q^\alpha$. Keeping in mind that $\Omega$ function
is invariant with respect to the Fourier transform, this conclusion
is also valid for the momentum space. Note that this kind of
discreteness depends on adelic quantum state of the universe. When
system is in an excited state, then the sharp discrete structure
disappears, and minisuperspace, as well as configuration space in
quantum mechanics, demonstrate usual properties of real space.

In spite of the very attractive features of the tachyonic
inflation, first of all, the rolling tachyon condensate, this
approach faces difficulties such as reheating (Sami, 2004). It
seems that both mechanisms, based on real tachyons - conventional
reheating mechanism and quantum mechanical particle production
during inflation - do not work. Recent results in nonlocal
($p$-adic) tachyon inflation (Barnaby et al., 2007; Joukovskaya,
2007), in which a $p$-adic tachyon drives a sufficiently long
period of inflation while rolling away from the maximum of its
potential deserve much more attention. The {\it classical}
$p$-adic models succeed with inflation where the real string
theory fails. In this paper we have calculated a {\it quantum}
propagator for the $p$-adic and adelic tachyons, found conditions
for the existence of the vacuum state of $p$-adic and adelic
tachyons, noted interesting relations with the minisuperspace
closed homogenous isotropic model in $(2+1)$ dimensions using
Einstein gravity with a cosmological constant and an antisymmetric
tensor field matter source (Djordjevic et al., 2002, Halliwell and
Myers, 1989). We have shown that the new results can give rise to
a better understanding of the $p$-adic and real quantum tachyons,
their relation via Freund-Witten formula and a possible role of
tachyon field as a dark matter. Our results can also be used as a
basis for further investigation of ($p$-adic) quantum mechanical
damped systems and corresponding wave functions of the universe in
the minisuperspace models based on the tachyonic matter with
different potentials. Further investigation should contribute to
the better understanding of quantum rolling tachyon scenario in a
real (Ambjorn and Janik 2004) and $p$-adic case.

\vskip 14pt {\large\textbf{Acknowledgements:}}

This work is partially supported by the Ministry of Science of the
Republic of Serbia under Grants 144014. Work on this paper is also
supported in part by the UNESCO-BRESCE/IBSP grants No. 875.854.7 and
No.875.922.8, within the framework of the Southeastern European
Network in Mathematical and Theoretical Physics (SEENET-MTP). We
would like to thank B. Dragovich, M. Hindmarsh, A. Linde, R. Kalosh,
S. Kar, A. S. Koshelev and A. Sen for helpful discussions.

\references

Ambjorn J. and Janik R. A. : 2004, \journal{Phys. Lett.},
\vol{B604}, 225.

Are`feva, I. Ya., Dragovich, B.,  Volovich, I. V. : 1988,
\journal{Phys. Lett.}, \vol{B200}, 512.

Aref'eva, I. Ya., Dragovich, B., Frampton, P. H., Volovich, I. V.
: 1991, \journal{Int. J. Mod. Phys.}, \vol{A6}, 4341.

Barnaby, N., Biswas, T., Cline, J. M. : 2007, \journal{JHEP},
\vol{0704}, 056.

Brekke, L., Freund, P. G. O. : 1993, \journal{Phys. Rep.},
\vol{233}, 1.

Choudhury, D., Ghoshal, D. : 2002, \journal{Phys. Lett.},
\vol{B544}, 231.

Dimitrijevic, D. D., Djordjevic, G. S., Nesic, Lj. : 2008,
\journal{AUC}, \vol{18}, 166.

Dimitrijevic, D. D, Djordjevic, G. S., Nesic, Lj. : 2008,
\journal{Fortschr. Phys.}, \vol{56}, No. 4-5, 412.

Dimitrijevic, D. D., Djordjevic, G. S., 2007, in The Sixth
International Conference of the Balkan Physical Union, edited by
Cetin, S.~A., Hikmet, I. (AIP, CP899), 359.

Dimitrijevic, D. D., Djordjevic, G. S., Nesic, Lj. : 2004,
\journal{Facta Universitatis, Series: Physics, Chemistry and
Technology}, \vol{3}, No 1, 7.

Djordjevic G. S., Dragovich B, Nesic Lj. D. and  Volovich I. V. :
2002, \journal{Int. J. Mod. Phys.}, \vol{A17}, 1413.

Djordjevic, G. S., Dragovich, B. : 2000, \journal{Mat.Theor.Phys},
\vol{124}, 1059.

Djordjevic, G. S., Dragovich, B., Nesic, Lj. : 1999, \journal{Mod.
Phys. Lett.}, \vol{A14}, 317.

Djordjevic, G. S., Dragovich, B. : 1997, \journal{Mod. Phys.
Lett.}, \vol{A12}, 1455.

Djordjevic, G. S., Dragovich, B.,  Nesic, Lj. : 2002,
\journal{Nucl. Phys. B Proc. Suppl.}, \vol{104}, 197.

Djordjevic, G. S., Dragovich, B.,  Nesic, Lj. : 2003,
\journal{IDAQP}, \vol{6}, 179

Djordjevic, G. S., Nesic, Lj. : 2005, \journal{Rom. J. Phys.},
\vol{50}, 289.

Djordjevic, G. S., Nesic, Lj. : 2005, in Mathematical, Theoretical
and Phenomenological Challenges Beyond the Standard Model:
Perspectives of the Balkan Collaborations, edited by Djordjevic,
G. S., Nesic Lj. and Wess, J, (World Scientific, Singapure) 197.

Djordjevic G. S., Nesic, Lj. : 2003, in Springer Lecture Notes in
Physics \vol{616}, edited by J. Trampetic and J. Wess (Springer,
Berlin/Heidelberg), 25 (hep-th/0412088).

Djordjevic, G.S., Dragovich B. : 2000, Proc. XII Conference on
Applied Mathematics, Igalo, Yugoslavia,  May 1998, Eds. D. Herceg.
K. Surla, N. Krejic,  Novi Sad 23.

Djordjevic, G. S., Dragovich, B. :  1997, \journal{Mod. Phys.
Lett.} \vol{A}, 1455.

Dragovich, B. : 1994, \journal{Theor. Mat. Phys.}, \vol{101}, 349.

Dragovich, B. : 1995, \journal{Int. J. Mod. Phys.}, \vol{A10},
2349.

Dragovich, B. : 1995, Proc. of the Third A. Friedmann Int. Seminar
on Gravitation and Cosmology, Friedmann Lab. Publishing, St.
Petersburg, 311.

Dragovich, B., Nesic, Lj. : 1999, \journal{Grav. Cosm.}, \vol{5},
222.

Dragovich, B. : 2009, \journal{Fortschr. Phys.}, \vol{57}, No.
5-7, 546.

Dragovich B., Khrennikov A. Yu., Kozyrev S. V., Volovich I. V. :
2009, \journal{Anal.Appl.}, \vol{1}, 1.

Dragovich, B., Nesic, Lj. : 1996, \journal{Facta Universitatis,
Series: Physics, Chemistry and Technology}, \vol{1}, No 3, 223.

Freund, P. G. O., Witten, E. : 1987, \journal{Phys. Lett.},
\vol{B199}, 191.

Garay, L. J. : 1995, \journal{Int. J. Mod. Phys.}, \vol{A10}, 145.

Garay L. J., Halliwell J. J. and Mena Maru\'gan G. A. : 1991,
\journal{Phys. Rev.} D43, 2572.

Garcia-Compean H., Obregon O. and Ramirez C. : 2002,
\journal{Phys. Rev. Lett.}, \vol{88}, 161301.

Gel'fand I. M., Graev I. M., Piatetskii-Shapiro I. I. : 1966,
\journal{Representation Theory and Automorphic Functions},
Saunders, London.

Gibbons, G. W. : 2003, \journal{Class. Quantum Grav.}, \vol{20},
321.

Halliwell, J. J.,  Myers, R. C. : 1989, \journal{Phys. Rev.}
\vol{D40}, 4011.

Halliwell J. J., Louko J. : 1989, \journal{Phys. Rev.} D39, 2206.

Hartle, J., Hawking, S. : 1983, \journal{Phys. Rev.}, \vol{28},
2960.

Jain, D., Das, A., Kar, S. : 2007, \journal{American Journal of
Physics}, \vol{75}, No. 3, 259.

Joukovskaya, L. : 2007, \journal{Phys. Rev.}, \vol{D76}, 105007.

Kar, S. : 2002, IITK preprint, \journal{A Simple Mechanical Analog
of the Field Theory of Tachyon Matter}, hep-th/0210108.

Khrennikov, A. : 1997, \journal{Non-Archimedean Analysis: Quantum
Paradoxes, Dynamical Systems and Biological Models}, (Kluwer Acad.
Publ., Dordrecht).

Lidsey, J. E. : 2007, \journal{Phys. Rev.}, \vol{D76}, 043511.

Perlmutter, S. et al. : 1999, \journal{Astrophys. J.}, \vol{517},
565.

Ruelle, Ph., Thiran, E., Verstegen, D., Weyers, J. : 1989,
\journal{J. Math. Phys.}, \vol{30}, 2854.

Sami, M., Chingangbam,   P., Qureshi,   T. : 2004,
\journal{Pramana}, \vol{62}, 765.

Sen A. : 2002, \journal{JHEP} 0204, 048; Sen A. : 2005,
\journal{Int. J. Mod. Phys.} {\bf A20}, 5513.

Schikhof, W. H. : 2006, \journal {Ultrametric calculus: An
introduction to p-adic analysis} Cambridge.

Tranberg,  A., Smit,  J., Hindmarsh,   M. : 2007, \journal{JHEP},
\vol{0701} 034.

Vladimirov, V. S., Volovich, I. V., Zelenov, E. I. : 1994, p-Adic
Analysis and Mathematical Physics, (World Scientific, Singapore).

Vladimirov, V. S., Volovich, I. V. : 1989, \journal{Commun. Math.
Phys.}, \vol{123}, 659.

Volovich, I. V. : 1987, CERN-TH, \vol{4781}.

Volovich, I. V. : 1987, \journal{Class. Quantum Grav.}, \vol{4},
L83.

Wiltshire, D. L. : 1996, in The Cosmology: Physics of the
Universe, edited by B. Robson et al., (World Scientific,
Singapore) 473.

\endreferences

\end{document}